\documentclass[a4paper]{jpconf}%
\usepackage{graphicx}
\usepackage{amsmath}

\usepackage{xspace}
\usepackage{xcolor}
\usepackage{siunitx}

\begin{document}
\title{Phasonic Diffusion and Self-confinement of Decagonal Quasicrystals in Hyperspace}

\author{Johannes Hielscher, Miriam Martinsons, Michael Schmiedeberg, Sebastian C.\@ Kapfer}

\address{Friedrich-Alexander University Erlangen-Nürnberg, Institute for Theoretical Physics I, Staudtstr.\@ 7, 91058 Erlangen}

\ead{johannes.hielscher@fau.de}%

\begin{abstract}%
We introduce a novel simulation method that is designed to explore fluctuations of the phasonic degrees of freedom in decagonal colloidal quasicrystals.
Specifically, we attain and characterise thermal equilibrium of the phason ensemble via Monte Carlo simulations with particle motions restricted to elementary phasonic flips.
We find that, at any temperature, the random tiling ensemble is strongly preferred over the minimum phason-strain quasicrystal.
Phasonic flips are the dominant carriers of diffusive mass transport in physical space.
Sub-diffusive transients suggest cooperative flip behaviour on short time scales.
In complementary space, particle mobility is geometrically restricted to a thin ring around the acceptance domain, resulting in self-confinement and persistent phasonic order.
\end{abstract}%

\section{Introduction}
Intrinsic quasicrystals on the colloidal length scale are of much interest, especially due to the accessibility of microscopic details. %
While self-assembly and phase behaviour (see \emph{e.\,g.\@} \cite{engel-2007, Martinsons_2019}), thermodynamics and phason elasticity \cite{Strandburg_1989PRL} have been studied for two-dimensional (2D) decagonal quasicrystals, the specific role of phasonic excitations is still elusive.
Phasonic degrees of freedom are unique to quasicrystals, visible as particle flips in physical space \cite{Socolar_1986, Kromer_2012}.
Simulations in continuous space \cite{Strandburg_1989PRL, Kiselev_2012} have found ``random tiling ensembles'' \cite{Henley_1988} %
with finite phason strain for various systems.
However, phasonic contributions cannot be isolated, and the vastly different time scales between phononic motions and phasonic flips remain a challenge for simulations.

Experimental studies on decagonal (3D axial) intermetallic quasicrystals do not detect a significant contribution of phasons to mass transport \cite{Khoukaz_2001}, rather suggest a regular vacancy mechanism for self-diffusion.
We present a simple hyperspace model for 2D colloidal quasicrystals without defects (dislocations, vacancies, surfaces), and can specifically identify phason-driven transport.
Flip simulations are in quantitative agreement with conventional Brownian Dynamics.

\section{Methods}
\begin{figure}[htbp]
\includegraphics{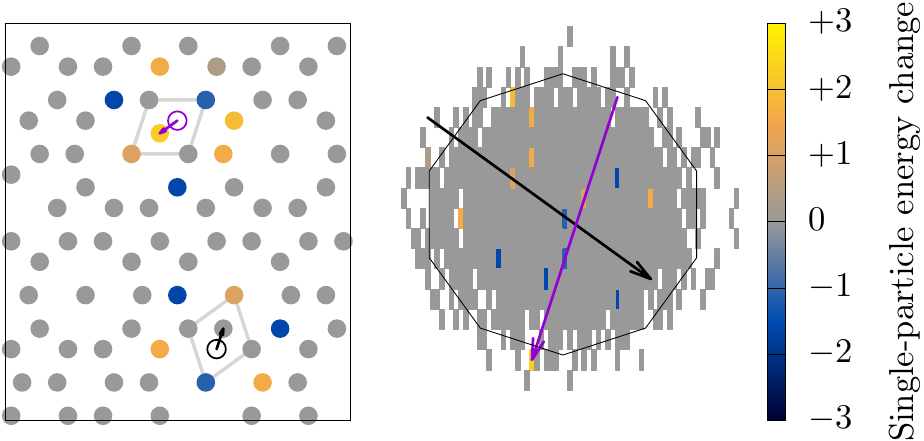}\hspace{2pc}
\begin{minipage}[b]{14pc}\caption{\label{fig_flipview}Two typical phasonic flips (arrows) moving a particle in physical (left) and complementary space (right; decagon: original acceptance domain). The energy difference (colours) of the flip originates from the adjacent particles in physical space.}
\end{minipage}
\end{figure}

We lift the particles of a 2D decagonal quasicrystal onto the 5-dimensional hypercubic integer lattice.
Moves of the particles are restricted to primitive hypercubic vectors which correspond to phasonic flips (see fig.~\ref{fig_flipview}).
Interactions are governed by the Lennard-Jones--Gauss pair potential $V_\textnormal{LJG}(r_\parallel)$ that is designed to support the two length scales of quasicrystalline structures by its two minima \cite{engel-2007, Hielscher--2017}.
The hyperlattice model
(absence of phonons)
allows for a restriction of distances to discrete values that were extracted from Brownian dynamics simulations \cite{Hielscher--2017}:

\medskip
\noindent%
\strut\hspace*{-6mm}\begin{minipage}[b]{0.6\textwidth}
\begin{equation*}
	V(r_\parallel)=\begin{cases}
		\num{-0.830} & r_\parallel=1/\tau\approx\num{0.618} \\
		\num{-1.949} & r_\parallel=1 \\
		\num{-0.328} & r_\parallel=\num{1.177} \\
		\infty & r_\parallel\leq\num{0.587}\quad\textnormal{(hard core)} \\
		0 & \textnormal{else,}
	\end{cases}
\end{equation*}
\end{minipage}%
\hspace*{-1mm}%
\begin{minipage}[t]{0.4\textwidth}
	\vspace*{-3.5\baselineskip}

	\noindent\includegraphics{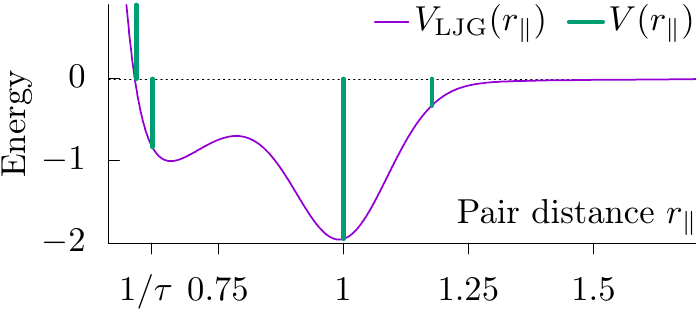}%
\end{minipage}%

\medskip
Simulations directly propose phasonic flips, accepted with the Metropolis Monte Carlo (MC) probability
$\exp(-(V_\textnormal{after}-V_\textnormal{before})/T)$
at some temperature $T=1/\beta$.
The simulation box is a periodic approximant to the %
decagonal quasicrystal with $N$ particles (here usually $N=1686$); the initial condition is the minimum phason-strain quasicrystal, given by the canonical (solid decagon) acceptance domain.
Brownian Dynamics simulations of the same system ($T=0.3$) were found to excite \num{6.5(24)} flips each $4\,\tau_\textnormal{B}$ Brownian time units.
In comparison, a MC sweep causes \num{6.4(25)} flips, hence corresponds to a physical time of $4\,\tau_\textnormal{B}$.
In terms of CPU time, our simulation accelerates phason flip dynamics by at least four orders of magnitude.

\noindent
The centre of mass performs a random walk $\langle\vec{r}_*\rangle(t)$ through hyperspace, where $*$ either denotes $\parallel$ for physical or $\perp$ for complementary space (see trajectory in fig.~\ref{fig_vetocloud}).
We examine mean-square displacements
$\langle \Delta r_*^2\rangle = 1/N \sum_{j=1}^N \left( \vec{r}_*^{(j)}(t)-\langle{\vec{r}_*}\rangle(t) -\vec{r}_*^{(j)}(0) \right)^2$
with a correction of the centre-of-mass motion.
Time-dependent diffusion coefficients %
$D_\parallel(t)=\langle r_\parallel^2(t)\rangle/(2t)$ approach a constant value $D_\parallel = \lim_{t\rightarrow\infty} D_\parallel(t)$ for diffusive transport.

\section{Results and Discussion}
\begin{figure}[ht]
\includegraphics{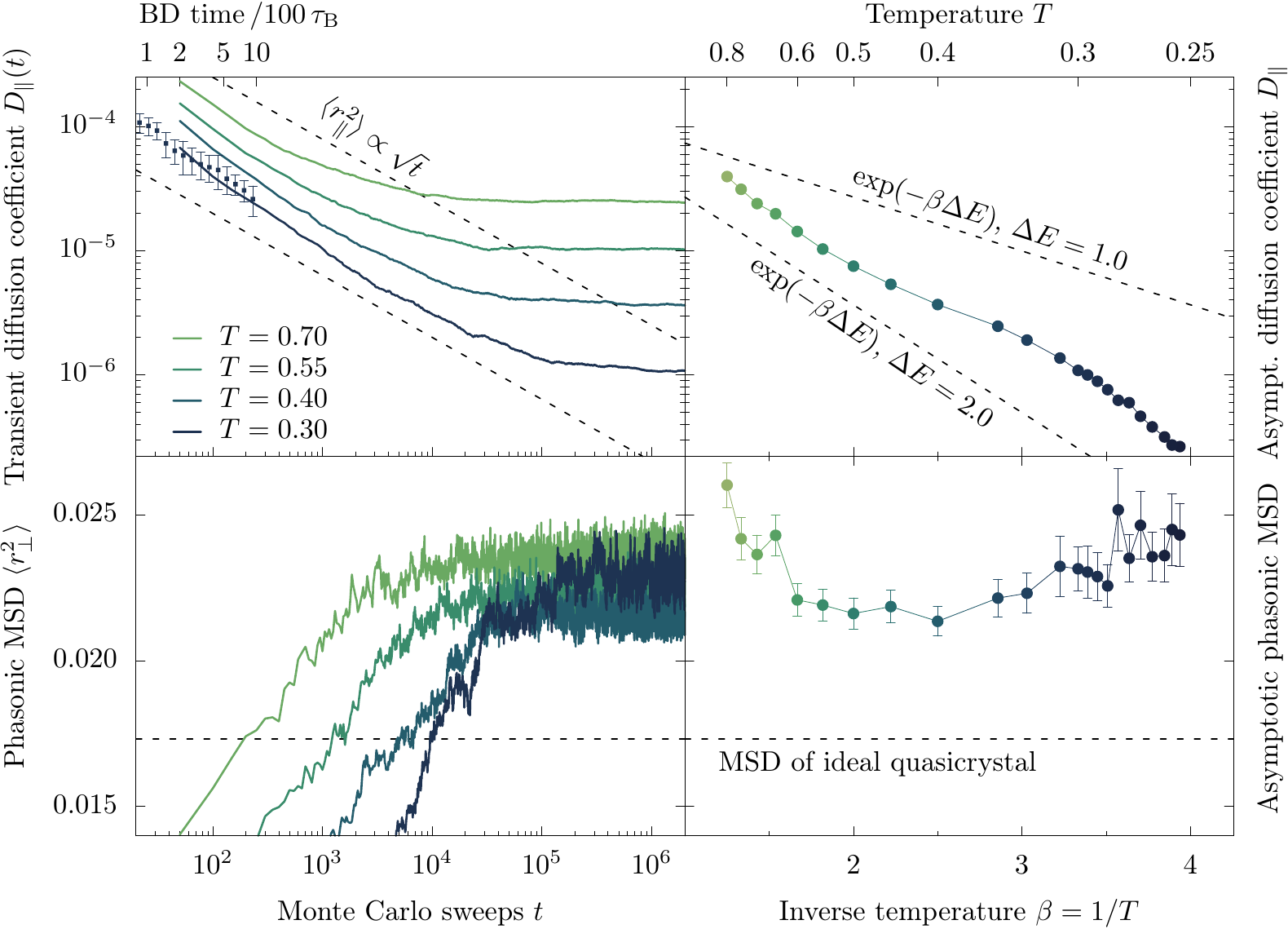}
\caption{\label{fig_msd}Transport in quasicrystals.
Left column: transients of Monte Carlo thermalisation (points: Brownian Dynamics data, scaled to $4\,\tau_\textnormal{B}$ per sweep); right column: asymptotic values versus temperature.
Top row: Diffusion in physical space is governed by thermal activation, with pronounced $\sqrt{t}$-like sub-diffusive transients.
Bottom row: In complementary space, the quasicrystal stays cohesive for arbitrarily long times, the mean-square displacement saturates without distinctive $T$ dependence.}
\end{figure}

Phason flip dynamics cause a rapid (less than \num[retain-unity-mantissa=false]{1e4} sweeps) build-up of \SI{20}{\percent} %
particles flipped from the initial quasicrystal.
Equilibration, \emph{i.\,e.\@} the saturation of time-dependent diffusion coefficients and the energy autocorrelation, takes less than \num[retain-unity-mantissa=false]{1e5} sweeps above $T=\num{0.3}$, and is very fast for $T>\num{0.5}$. %
The data of fig.~\ref{fig_msd} is recorded after %
initial thermalisation of \num[retain-unity-mantissa=false]{1e5} sweeps.
We notice some flips that decrease the total energy. %

\subsection{(Sub)diffusive transport in physical space}
Transport, seen from physical space, asymptotically becomes diffusive.
The diffusion constant $D_\parallel$ (fig.~\ref{fig_msd}, top right) depends exponentially on $T$, with an activation barrier $\Delta E$ in the order of energetic costs of an individual flip.
This thermal activation reminds of vacancy diffusion in crystalline solids, that relies on a finite density of point defects (vacancies).
However, the transport in colloidal quasicrystals is carried by phasonic flips, common to all (even defect-free) systems.
Transient transport is sub-diffusive with approximate $\langle r_\parallel^2\rangle \propto \sqrt{t}$ over several decades in time. %
This anomalous exponent is known for single-file diffusion in 1D systems \cite{Levitt_1973, Kollmann_2003}, where particles are blocking mutual passage.
Similarly, the strong geometrical interlock of phasonic flips imposes more severe constraints on motion than expected from the 2D nature of the system (\emph{cf.} blocking in complementary space, see fig.~\ref{fig_vetocloud}), prolonging the approach to diffusive asymptotics.

\subsection{Complementary space: Self-confinement}
\begin{figure}[htbp]
\includegraphics{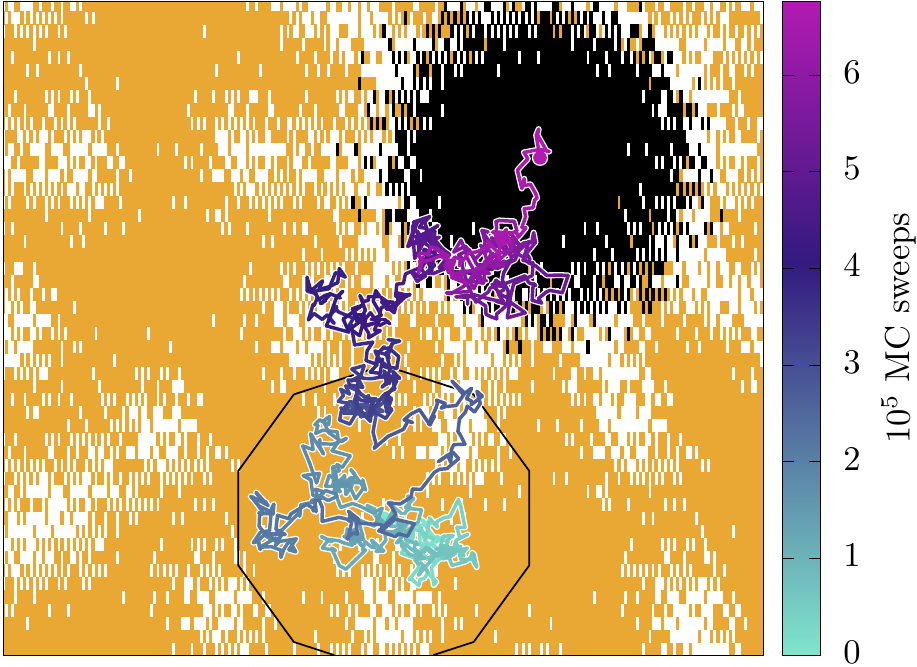}\hspace{1pc}
\begin{minipage}[b]{14.5pc}\caption{\label{fig_vetocloud}The quasicrystal drifts through complementary space ($T=0.40$, coloured line: trajectory of the centroid). Solid black: current particle positions; decagon: acceptance domain (initial occupations). Each orange hyperlattice site lies inside the hard core of some particle in physical space, so that only the white sites are possible as destinations of flips (self-confinement).}
\end{minipage}
\end{figure}
The projection into complementary space (black boxes, fig.~\ref{fig_vetocloud}) reveals the maintenance of cohesion over arbitrarily long times, at any examined temperature.
During warm-up, flips to outside the solid core of the quasicrystal
rapidly establish a halo of fractional occupation.
Its width is quantified by the rise of $\langle r_\perp^2\rangle$ above the mean-square displacement of the perfect quasicrystal. %
The saturation of $\langle r_\perp^2\rangle$ indicates motion in a restricted area, %
\emph{i.\,e.\@} geometrical self-confinement within a pore in complementary space:
hyperlattice sites further away than the pore radius are mostly blocked by the hard cores in physical space (shaded areas in fig.~\ref{fig_vetocloud}).

\section{Conclusions}
We have studied the phasonic equilibrium of decagonal quasicrystals governed by a short-ranged pair potential.
A simulation model in hyperspace exclusively treats phasonic degrees of freedom via explicit elementary flips.
Comparison with Brownian Dynamics estimates each Monte Carlo sweep to be equivalent to the physical time of about $4\,\tau_\textnormal{B}$ Brownian times.
This emphasises the efficiency of phasonic thermalisation, and enables long-term studies of phason dynamics.

The perfect quasicrystal (minimum phason strain) is far from equilibrium.
Rather, equilibrium is distinguished by the presence of phasonic excitations, with amplitudes that hardly depend on temperature.
The asymptotic transport in physical space is diffusive with thermally activated diffusion constants.
Though similar to vacancy diffusion in periodic crystals, it is driven by the intrinsic degrees of freedom of the quasicrystal.

In complementary space, the quasicrystal forms a phasonic halo and stays confined as a whole, stabilised by self-imposed energetic and geometric constraints on phasonic flips.
The dynamics is confined to particle motions inside the halo, apart from to the (unbound) drift of the centre of mass.

\ack
We acknowledge support by the Deutsche Forschungsgemeinschaft (DFG)
via grant Schm 2657/4 and the Research Unit \emph{Geometry and Physics of Spatial Random Systems} (grant Me 1361/12).

\end{document}